\documentclass[10pt, conference]{IEEEtran}
\usepackage{times}  
\usepackage{subcaption}
\usepackage{color,soul}
\usepackage{xcolor}

\usepackage{url}
\usepackage{tikz}

\begin{document}




\title{Infinity: A Scalable Infrastructure for In-Network Applications}

\author{\IEEEauthorblockN{Marcelo Abranches}
\IEEEauthorblockA{
\textit{University of Colorado - Boulder}\\
Boulder, USA \\
marcelo.abranches@colorado.edu}
\and
\IEEEauthorblockN{Karl Olson}
\IEEEauthorblockA{
\textit{University of Colorado - Boulder}\\
Boulder, USA \\
karl.olson@colorado.edu}
\and
\IEEEauthorblockN{Eric Keller}
\IEEEauthorblockA{
\textit{University of Colorado - Boulder}\\
Boulder, USA \\
eric.keller@colorado.edu}}

\maketitle
\begin{abstract}

Network programmability is an area of research both defined by its potential and its current limitations. While programmable hardware enables customization of device operation, tailoring processing to finely tuned objectives, limited resources stifle much of the capability and scalability desired for future technologies. Current solutions to overcome these limitations simply shift the problem, temporarily offloading memory needs or processing to other systems while incurring both round-trip time and complexity costs. To overcome these unnecessary costs, we introduce Infinity, a resource dis-aggregation method to move processing to capable devices while continuing to forward as the original owner, limiting unnecessary buffering and round-trip processing. By forwarding both the processing need and associated data simultaneously we are able to scale operation with minimal overhead and delay, improving both capability and performance objectives for in-network processing.
\end{abstract}
\section{Introduction}


The Internet was designed over 40 years ago, and in the decades that followed, the ubiquity of it both served as a great testament to its architecture, but also introduced new operational challenges.  In particular, with new applications and uses there were new requirements, many of which were not supported under the current architecture.  While the research community followed with innovative solutions, the ossification served as an impediment to adoption~\cite{ossification_hotnets04}.  Today, this problem is even more pronounced with the rise of edge computing and proposals for tactile applications that have the potential to revolutionize health-care, for example.  These applications require high availability, security, and low latency that the current Internet cannot provide.


Towards the goal of enabling this future, researchers have turned to technology ranging from virtualization~\cite{etsinfv, ossification_hotnets04} to programmable network hardware~\cite{openflow, p4, forwardingmetamorphosis}.  Our focus here is on the programmable network hardware. Traditional networking equipment follows a fixed function design model (e.g., ASICs) typically optimized for a narrowly defined role, such as a data center, distribution, or core device. This design  approach  provides  a  refined  high  performance architecture with limited versatility outside predefined operational  scopes.  However,  this  limits  the  degree  of  potential innovation,  requiring  operators  to  implement  solutions  that align  with  often  out-dated  pre-coded  characteristics  of  the network device.  This changed with the introduction of the Protocol Independent Switch Architecture (PISA)~\cite{forwardingmetamorphosis}, which provides a hardware design in which the protocol parsing and packet processing are entirely programmable.   


What this means for the next-generation Internet is that there are now unique opportunities for in-network computing capabilities at high-performance and low-latency that were not possible before.  In recent years, in-network applications have been used to reduce latency for key-value stores~\cite{netcache}, support distributed locking~\cite{netlock}, and performing data aggregation to accelerate machine learning~\cite{switchml}.  These all point to the great potential for in-network computing that comes with these new programmable hardware architectures.


Unfortunately, there's a catch.  The architecture is characterized by a pipeline of match-action tables, with, for example, an associate TCAM (ternary content-addressable memory) to perform quick lookups, an arithmetic-logic unit (ALU) to enable flexible actions, and Static random-access memory (SRAM) to enable state storage.  Being that it is hardware, there's a limitation on the amount of each that are available.  At the same time, innovations seek to do more in-network - both in terms of the complexity of the designs, but also in the number of concurrent applications we want to support.  Interestingly, general purpose computing platforms (CPUs) have similar resource limitations, but have established operating system abstractions that mask them (e.g., virtual memory for extended memory, and CPU scheduling for multiple concurrent applications).  We argue that programmable hardware needs its own set of abstractions to meet the growing needs of in-network applications and dynamic architectures.

One such related work that sought to overcome the resource constraint around memory on programmable switches, was TEA (for Table Extension Architecture)~\cite{tea}.  For this, they built on recent advances in computing infrastructures that leveraged remote direct memory access (RDMA) capabilities to enable fast remote memory~\cite{farm, infiniswap}. By using RDMA to a remote server, they illustrated how they could implement enough of the RDMA protocol in a P4-enabled switch to be able to access memory on a remote server. This meant that if they ran out of memory on a switch, they could buffer a packet and do a remote read to a data store, effectively extending the memory available.  Unfortunately, this has three key limitations: (i) it only works for memory resources, (ii) it incurs a buffering penalty on the switch, and (iii) it incurs a round-trip time in the middle of the packet processing.  



To address these three challenges we propose \textit{Infinity}, a programmable network architecture which abstracts networking hardware into virtual aggregated hardware sets, giving in-network applications the illusion of having infinity processing capacity. Infinity ensures that in-network applications can locate resources within the aggregated hardware sets necessary to meet processing objectives by leveraging data plane dis-aggregation, scale-out techniques (sequential decomposition/vertical scaling and scale-out horizontal scaling), and per-function tailored performance requirements. 

With Infinity we enable the following contributions to improve current in-network applications by leveraging programmable networking hardware:
\begin{itemize}
    \item We provide abstractions for building scalable and flexible in-network applications on top of programmable hardware enabling high-performance and low-latency processing.
    \item We enable in-network applications to be deployed on top of our \textit{Virtual Infinity Switch} abstraction, giving them the illusion of infinite computation resources.
    \item We provide a mechanism to fully and efficiently utilize the network infrastructure within the data center.
\end{itemize}

The remainder of this paper is organized as follows: In section \ref{sec:motivation} we frame the problem of fully flexible internet architectures by highlighting limitations with current programmable solutions. We then introduce our architectural solution, Infinity, and demonstrate how we address the limitations of current dis-aggregation and scalability solutions in section \ref{sec:solution} before providing examples of use cases in section \ref{sec:use_cases}. In section \ref{sec:discussion} we propose future directions with Infinity to enable fully flexible architectures while also addressing limitations to our approach. Finally, in section \ref{sec:conclusion} we conclude the paper and discuss the next steps in the direction of realizing our vision.
\section{Programmable Network Hardware Resource Limitations and Disaggregation Challenges}
\label{sec:motivation}
\subsection{Programmable Hardware}
Modern networks operate in a highly dynamic environment of competing stakeholder interests. Here, architects are pressured to provide solutions which support dynamic per-individual use-cases with high performance, efficient, and agile functionality. Competitiveness of a provider is therefore linked to how well they can provide dynamic solutions that are naturally at odds with current static architectures. The versatility provided by programmable hardware therefore appeals to the modern network, enabling tailored  functionality to meet performance, flexibility, energy efficiency, or cost effective business objectives.

With programmable networking hardware operators can add new functionality within the network, bypassing traditional development pipelines while increasing product life-cycles. Two common examples of programmable hardware include: protocol independent switch architecture (PISA) based switches and programmable SmartNICs. 


To allow for simple configuration, domain specific languages (DSLs) like P4~\cite{p4} were introduced to bridge the gap between low-level hardware configuration and operator knowledge. Critically, P4 is vendor agnostic with platform specific compilers provided by all of the common programmable hardware manufacturers~\cite{agilio, tofino, p4fpga}. P4 has constructs capable of programming the different components of a programmable networking device, producing both a flexible and feature complete data plane. For example, with P4, an operator may tailor their hardware to enable custom packet header parsers and de-parsers, or to define matching fields and actions to be included on the match action tables (MATs). It also provides a runtime, that allows defining application programming interfaces (APIs) for data/control plane communication.



\subsection{Resource Limitations with Programmable Hardware}

Although both SmartNICs and programmable switches offer significant customization with packet processing capabilities, they have finite resources which impose constraints on in-network processing scalability and desired functionalities. For example, modern programmable switches are limited in fast memory resources (i.e., SRAM) \cite{silkroad, tea} and have a finite amount of processing units (e.g., ALUs) \cite{acceleratedsc} that can be used to compose the processing pipelines. This limitation creates a challenge to run in-network applications with complex logic while simultaneously maintaining high-performance and scalability~\cite{mantis}. Adding to this challenge, diverse programmable hardware sets are built with different underlying architectures and targeted functionalities, making no single type of equipment suitable for running all problem sets~\cite{flightplan}.

\subsection{Hardware Abstractions}

In order to overcome limitations with SRAM availability on programmable switches, a table extension architecture (TEA) \cite{tea} was developed to leverage remote memory on RDMA capable commodity servers. Utilizing RDMA effectively works to expand the memory capacity on programmable hardware, making them more suitable to run stateful NFs at scale (e.g., network address translators (NATs), stateful load balancers, and firewalls). This approach also creates an abstraction layer to extend the switches' MATs (originally stored in local SRAM to ensure high-performance) using remote DRAM, which is accessed directly from the switches data plane whenever an entry cannot be found within the local SRAM. To avoid stalling packet processing during remote look ups, TEA further provides a method in which a triggered packet for the remote look-up is encapsulated in the RDMA request and temporally stored on remote DRAM. When the RDMA server sends the response to the look up, the packet is retrieved along with remote state contents, after which the deferred packet is processed according to the contents retrieved from remote memory. TEA further leverages the RDMA protocol to minimize overheads that could impact its data plane performance. For example, a single set of RDMA state for each queue pair (QP) is maintained based on assumptions about the underlying Ethernet network topology (i.e., Ethernet is considered to be reliable and switches are considered to be directly connected to the RDMA hosts).

\subsection{Challenges with Disaggregation}
\label{subsec:challenges}


While TEA can scale hardware resources on a programmable switch through memory dis-aggregation, deferring packet processing during state look ups increases processing inefficiencies on the network. First, packets will suffer from added latency caused by at least one round-trip time (RTT) before they are finally processed. Second, sending the packet to remote data store and querying state on remote memory will cause extra load on the network links. Third, while this approach can increase available memory for the programmable hardware entities, it cannot scale processing resources, as the packets will be returned to the same entity for further processing. A final further limitation of this approach is that TEA leverages precious DRAM resources on remote hosts which could instead support applications to run business logic (e.g., big data, analytics, etc).

\section{Introducing Infinity}
\label{sec:solution}

While programmable network hardware provides exciting avenues for introducing in-network applications, there's an inherent limitation in that the devices themselves are resource-constrained. 
To enable a new level of performance, scalability and flexibility for in-network applications, while ensuring efficient usage of the underlying infrastructure, we present \textit{Infinity}, a system that can virtualize multiple programmable networking hardware components into virtual aggregated hardware sets, giving applications the illusion of having infinite processing and memory capacity. To do so, with Infinity we introduce a set of networking hardware composing primitives which enable operating system like abstractions to dynamically scale the resource allocations for in-network applications. 

Before discussing Infinity, it is useful to understand the typical flow for programmable switches today.  Traditionally, in-network applications are written in a domain specific language like P4, which is first compiled and then mapped to finite physical resources on a switch based on an architecture description of the target. This target description will indicate the amount of SRAM, number of ALUs, etc. available.  Within the P4 program, the developer's program has an impact on the overall resource usage - e.g., it has to specify the size of each table statically at compile time.  If a P4 program is successfully compiled for a given hardware target, it can run at line rate, and the compiler will use resources as needed within the constraints of the target.  

\begin{figure}
\centering 
\captionsetup{font=small}
\includegraphics[trim={3cm 2cm 2cm 3cm}, clip, scale=0.4]{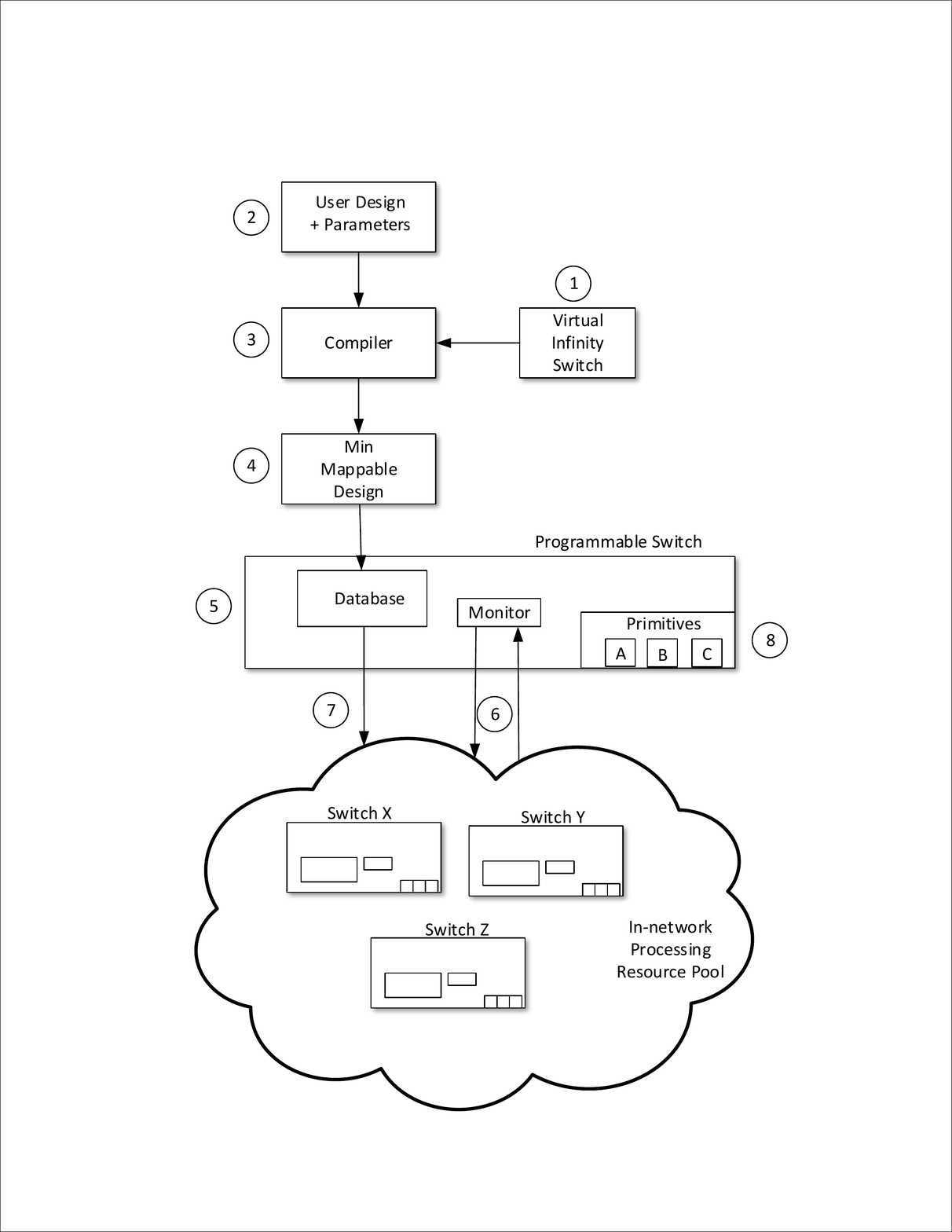}
 \caption{\textbf{Infinity - High Level Deployment} (1), (2) Infinitiy virtual switch architecture is integrated with custom user defined parameters via P4 Language. (3) Compiler prepares system resulting in minimal mappable architecture and loaded onto programmable switch (4). (5) Switch runs Infinity architecture collecting in-network resource metrics from participating systems (6). (7) When action requires resources beyond capability of host, Infinity references resource pool and (8) employs one of its scaling primitives (subsection \ref{subsec:primitives}) as needed.}\label{fig:protocol}
\label{fig:fig1}
 \vspace{-2em}
\end{figure}

In contrast, with Infinity, we aim to perform more of the resource mapping at run-time, such that we can provide the abstraction of an unlimited set of resources.  Figure \ref{fig:fig1} shows a high level representation of Infinity's design.  Infinity introduces a compiler that has as a \textit{Virtual Infinity Switch} (VIS) as a target. The compiler is responsible for generating a minimal design for a given in-network application that can be mapped into a VIS.  By minimal design, we mean the smallest unit that could be instantiated - e.g., continuing the example above of having to specify the table sizes, we can now assume a reasonable table size and know that it can be expanded later.  Further, the VIS abstraction enables an in-network application to expand the resource boundaries of a single switch, ensuring that the developer does not need to care about constraining the processing logic to fit into a single bounded entity. 

With a viable minimally mapped design the operator can, at load time, provide that design to the Infinity controller which will in turn place the application onto available resources.  Infinity will then continue to monitor the deployed in-network applications and identify bottlenecks, such as SRAM or TCAM exhaustion, due to too many flows being processed. Once a bottleneck is identified, Infinity will employ one of its primitives to expand available resources for the in-network application, as we will demonstrate next.

\subsection{Virtual Infinity Switch}
\label{subsec:vis}
Infinity's main goal is to seamlessly extend resources for running high-demand in-network applications. In this direction, we extend the concepts of the Single Big Router (applied to routers) \cite{keller_paas} and Single Big Switch (applied to OpenVswitch) \cite{bigswitch, ovs} to introduce VIS, an idealized abstraction of a programmable data plane with infinite resources (infinitely many pipelines, infinitely many stages, infinite memory, etc.).

After compilation, in-network application are mapped to a pipeline of programmable forwarding engine stages (or other processing elements) that implement the desired logic. Typically, these map to a single device, which is resource constrained and may prevent applications from being deployed if it's processing requirements go beyond the capacity of the device. This limitation may also prevent in-network applications from continuing to operate when facing resource exhaustion.

To enable a set of switches to appear as single entities with virtually infinity resources, each physical switch inside a VIS has a base functionality which supports an overlay network that encapsulates the original packets and transparently forwards them to a destination switch for further processing as needed. The overlays are built on top of a light-weight forwarding logic on each switch (i.e., custom header parsing, lookup ID, and forwarding). This enables flexible composition of primitives that support scaling hardware resources to meet application demands. 


Each switch is also assumed to be partially reconfigurable. That is, the switch does not require the entire programmable hardware to be flashed in its entirely or to be taken offline during data plane reconfiguration. In this manner, Infinity can dynamically allocate processing resources on the switches without disrupting the currently deployed packet processing applications. While this is not entirely supported on some hardware targets for P4, it is supported on FPGA targets which do support partial reconfiguration, and it is inevitable (in our opinion) for ASIC based switches to support P4 as targets.

\subsection{Infinity Primitives}
\label{subsec:primitives}

With the VIS abstraction, the underlying network of switches become a pool of resources for a controller to optimize the mapping.  To do so, the VIS abstraction needs a collection of primitives that support a dynamically scaling a design. For this, Infinity provides a set of primitives to enable meeting demands with different requirements. The primitives are designed to enable flexible composition of hardware elements allowing scaling resources for in-network applications as needed. 


\textbf{Primitive 1 - Sequential Decomposition:} The first primitive enables a processing pipeline to be split at any point and implemented across two or more switches. This enables applications to dynamically increase processing stages, allowing for logic which expands the processing capabilities of a single switch (i.e., applications can use more physical pipeline stages than is available on a single switch).

\begin{figure}
\centering 
\captionsetup{font=small}
\includegraphics[trim={2.5cm 18cm 1cm 3cm}, clip, scale=0.45]{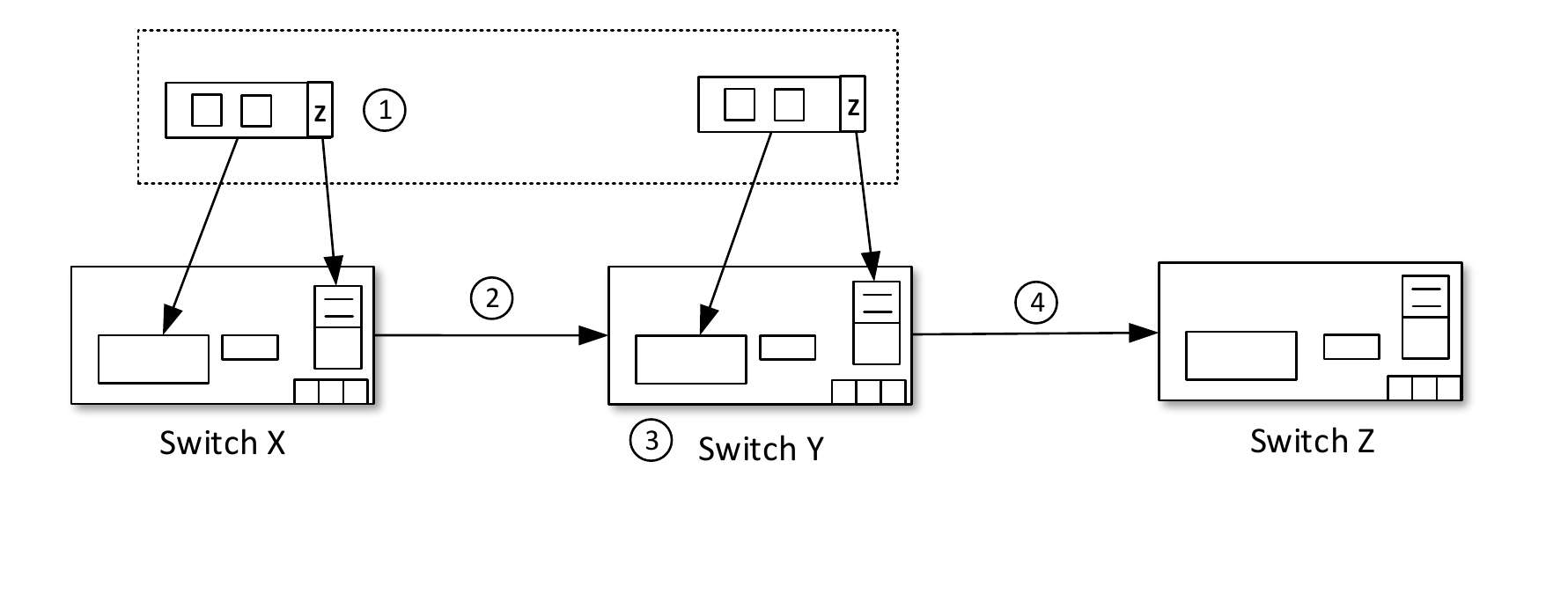}
 \caption{\textbf{Infinity - Sequential Decomposition Operation} (1) The controller determines to sequentially decompose the pipeline and place the first segment on Switch X. The controller sends to Switch X that packets are to be encapsulated with the target header and forwarded according to the lookup table that was built as the overlay (2). (3) Intermediate devices forwards according to packet header until it reaches in-network processing host (4), which decapsulates and processes with available resources. 
 }
\label{fig:fig2}
 \vspace{-1em}
\end{figure}

To realize this, we require a mechanism to connect the segments of the pipeline spanning multiple switches.  Figure \ref{fig:fig2} shows how Infinity realizes this primitive. Here, a given switch on the pipeline processes a packet and at the point the cut was made (to make more physical pipeline stages available), Infinity will insert logic to encapsulate it with a custom tag that simply has an ID of the switch where the next segment is mapped to. This is where the pre-configured overlay comes into play - each switch has forwarding logic that can lookup, based on that tag, and forward the packet. When the packet reaches the target switch, the packet is de-encapsulated and processed at the next segment.


\textbf{Primitive 2 - Horizontal Scaling:} The second primitive is horizontal scaling.  Similar to computing (where there can be replica instances of an application), this primitive enables scaling-out resources in order to increase bandwidth or the number of supported flows (as one example) for a given application. Horizontal scaling is allowed at two levels: the pipeline, where the full processing pipeline is replicated on another switch, or pipeline segment level where a pipeline is partially replicated on another switch.

\begin{figure}
\centering 
\captionsetup{font=small}
\includegraphics[trim={2cm 19cm 1cm .25cm}, clip, scale=0.45]{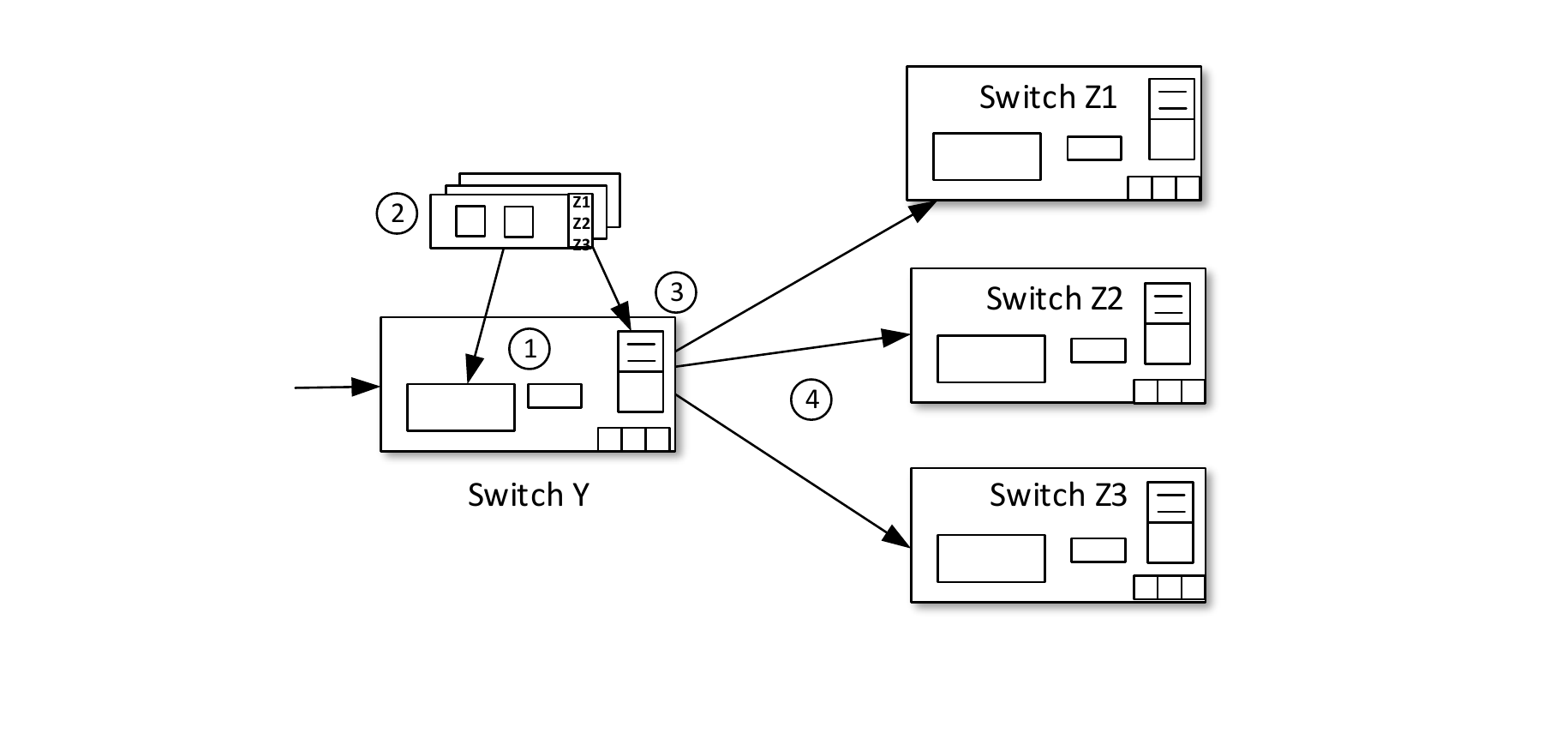}
 \caption{\textbf{Infinity - Horizontal Scaling Operation} The Infinity controller determines there's a need to horizontally scale, placing replicas on switch Z1, Z2, and Z3.  (1)Extra logic is added to Switch Y (the preceding segment), which will load balance across these three.  (2)The controller sends the switch ID for Z1, Z2, Z3 to Switch Y. (3)This forms the encapsulation that happens at Switch Y, and then packets are forwarded towards the replicas (4).  When the packet arrives at a replica, it decapsulates the tag and processes the packet.}
\label{fig:fig3}
 \vspace{-2em}
\end{figure}

To realize this, we need to insert extra logic to connect to the replicas, as shown in Figure \ref{fig:fig3}. First a pipeline (or segment) is replicated to another switch. Then, a load balancing element in the pipeline appends the segment preceding the horizontally scaled segment.  If it's the first segment or entire pipeline, this would be at the entry point of the network.  One complicating factor is that Infinity needs to know how to split the traffic - what traffic should go to each replica.  For this, the application needs to specify a key upon which to partition any memory resources.  For example, the key can be the 5-tuple of a flow to ensure all packets for the same flow are processed by the same replica.  But, a 5-tuple is not the only way to partition traffic, so we leave this to the application.


A second complicating factor is that the controller needs to know when resources are being exhausted.  For logic resources, this can be determined by the bandwidth, but for memory, this is a bit more challenging. To support this awareness, we require the applications to insert an extra bit into tables they want to be expanded. This bit would represent marking space as 'used' or 'unused'.  For example, in a dynamic NAT, as new entries are added, they will set the bit to used, and when they time-out or the flow ends, the bit will be set to unused.  This allows us to monitor for resource exhaustion.

\textbf{Primitive 3 - Vertical Scaling:} For the final primitive, Infinity supports vertical scaling.  In contrast to horizontal scaling, which adds extra replicas, vertical scaling makes individual instances (which don't have to be replicas) bigger - we allocate more resources to it.  As an analogy, consider a virtual machine with 1GB of memory allocated to it.  Horizontal scaling will launch an identical virtual machine with 1GB of memory, whereas Vertical scaling will change the allocation of the virtual machine to be 2GB.  Vertical scaling can be desirable, for example, in a case where it is not possible to use the horizontal scaling primitive, because it is not possible to use keys to efficiently balance traffic among replicated instances, or if there is no available switch resources to replicate a pipeline or a segment. 

There are two ways this can be realized in programmable switches. This first is through disaggregation. This is the mechanism introduced in TEA, where as the switch runs out of memory resources, it can buffer the packet, send a lookup request to remote memory, and then when the lookup returns, it will continue processing as if it got that state from local memory.  This is practical due to the speed of RDMA, and through keeping more used state within local memory. 

A second mechanism is through migration.  As an example, if the application cannot afford the performance penalty and we desire all memory to be local, but local resource constraints cannot fulfill the requirement, we must move it to a different switch (with a larger allocation).  This would then require copying over the state as part of the migration, which can be done in a live manner~\cite{lime}.  

\subsection{Orchestrating Infinity}
\label{subsec:orchestrating}
To enable its operation, Infinity relies on a compiler, which maps how physical resources are allocated for a given in-network application. Infinity also relies on a controller with a global view and influence over the network, enabling dynamic resource scaling by leveraging the Infinity primitives.

\textbf{Infinity Compiler:} As we can see in figure \ref{fig:fig1} Infinity provides a target model with an abstract description of the virtual infinite switch (VIS). This provides information regarding the type of hardware elements that compose the VIS, such that the compiler can map to it.  But, as mentioned, as resources are dynamically expandable, the output of the compilation is a minimally mapped design that consists of the unit of deployment from which the application can be expanded at run-time.

\textbf{Infinity Controller:} 
It is the job of the Infinity controller to find free resources within the pool of network switches and generate a mapping which can be deployed on the physical target. To do this, once the minimally mapped design is placed (allocated to a switch and loaded), the Infinity Controller monitors the current level of utilization for the critical hardware elements that may impact performance. To detect hot spots, Infinity leverages telemetry capabilities on switches and also functionalities implemented on the data plane. For example, as mentioned, SRAM usage on each hardware element is controlled by having each application to update a usage flag once a new flow is added. If a hot spot is detected the controller will work to mitigate it by using one or more of the scaling primitives described in \ref{subsec:primitives}. For example, if an application runs out of SRAM it can leverage the horizontal scaling primitive to increase capacity and process new flows. In this case the controller will replicate the affected pipeline on another switch with enough available resources, and will update the load-balancing rules, ensuring traffic is split among the parallel instances based on a given partition key (e.g., 5-tuple). This enables Infinity to act on a feedback loop, ensuring that applications will meet defined \textit{service level objectives} (SLOs). 



\section{Use Cases}
\label{sec:use_cases}

Pulling it all together, in this section we illustrate a couple of simple examples that illustrate how Infinity could be used to deploy in-network application that can scale to automatically expand the resources per the application needs.  



\textbf{Layer 4 Load Balancer:} A layer 4 load balancer's main purpose is to serve as an entry point for a scalable service, directing traffic to different servers to efficiently distribute the load on each.  One of the important characteristics in load distribution is awareness for flow affinity, such that all traffic in a TCP session, for example, is directed to the same server.  As such, the load balancer, which can be implemented in modern, programmable switches, needs to store state on the switch to remember which server was chosen for each flow.  This means that the amount of SRAM allocated is the limiting resource - if you allocate too much, you're wasting valuable resources on the switch, allocate too little and you may not be able to support enough flows.  With Infinity, this concern is alleviated with the primitives to scale out.  Here, lets say we end up using horizontal scaling.  In this case, the 5-tuple would be specified as the partitioning key, and there would be extra logic that splits traffic between two switches to perform the layer 4 load balancing service.   


\textbf{Access Control Lists:}
An access control list is a set of rules that define whether certain traffic should be permitted or denied passage beyond this particular network point.  It would consist of some classification and filtering logic that's coupled to a table with the set of rules.  The rules would be added dynamically by a control or management plane application.  With this, memory is again, a limiting resource, and directly determines how many rules that could be supported.  Ideally, we want to support as many as the operator wishes.  As such, we need scaling.  If we assume vertical scaling, perhaps because there's wild carding in the rule sets, then there is no clear way to partition traffic.  This is ok since these lists are likely change infrequently. In this case, when we detect that the process is running out of memory on the switch, we can initiate a vertical scaling operation to just allocate, for example, 25 percent more memory than is currently allocated.


\section{Discussion and Future Work}
\label{sec:discussion}
In the previous sections we saw the benefits that we expect from our proposal. Currently we are designing Infinity and there are some open questions and unexplored opportunities that we should address in order to fully realize our system.

\textbf{How to manage scaling decisions:}
To allow applications to fully benefit from Infinity, we need to have clear mechanisms to guide selection of the most appropriate scaling primitives (\ref{subsec:primitives} for a given scenario. For example, we need to decide if the allowed scaling primitives for a certain application should be decided and implemented by the programmer, or should another mechanism enable Infinity to automatically decide the ideal primitive when faced with a resource contention scenario? We aim to answer this question in a future work.


\textbf{Meeting SLOs:}
Redirecting flows for processing by another entity may introduce latency to the NFs. We argue that different applications have different latency requirements, a premise that Infinity can leverage to select prioritized flows for processing by the local NF while only redirecting lower priority traffic to remote NF instances. This would ideally avoid overloads on hardware components while maintaining SLOs.

\textbf{Leveraging SmartNICs:} In-network application SmartNICs can also benefit from the VIS abstraction. We plan as future work to extend Infinity and enable SmartNICs to participate on the VIS abstraction. This has the benefit of allowing further extension of resources for running high-demand in-network applications. For example, once a cryptography accelerator on a SmartNIC starts to suffer from queue build-up, Infinity can instruct the NF leveraging this feature to start sending a portion of its flows to an idle SmartNIC inside its VIS by leveraging the horizontal scaling abstraction. Another benefit of this extension will be to add SmartNICs processing capabilities to a hybrid VIS composed by both switches and SmartNICs, allowing for an optimal infrastructure utilization design.




\section{Conclusion}
\label{sec:conclusion}
In this vision paper we described Infinity, a system that allows in-network applications to be deployed on top of a programmable switch fabric with virtually infinite resources. We see Infinity as an important step towards the next-generation dynamic network; an architecture to support new in-network applications while enabling increased performance, scalability and flexibility that current static solutions cannot provide.

\section*{Acknowledgments}
This work was supported in part by NSF Grants 1652698 (CAREER) and the Coordenação de Aperfeiçoamento de Pessoal de Nível Superior - Brasil (CAPES) - Finance Code 001.

\bibliographystyle{abbrv} 
\begin{small}
\bibliography{bibliography}
\end{small}

\end{document}